\documentclass{cs19proc}

\editors{G.~A. Feiden}
\publisher{Zenodo}
\conference{The 19th Cambridge Workshop on Cool Stars, Stellar Systems, and the Sun}
\conferencedate{2016}

\title{Superflare G and K Stars and the Lithium abundance}
\author{Maria M. Katsova,$^{1}$ 
        Moisey A. Livshits,$^{2}$
	Tamara V. Mishenina,$^{3}$
	Bulat A. Nizamov$^{1, 4}$
	}

\affiliation{$^{1}$ Sternberg State Astronomical Institute, Lomonosov Moscow State University, Moscow, Russia\\
	$^{2}$ Pushkov Institute of Terrestrial Magnetism, Ionosphere and Radio Wave Propagation of Russian Academy of Sciences, Troitsk, Russia\\
	$^{3}$ Astronomical Observatory, Odessa National University, Odessa, Ukraine\\
	$^{4}$ Department of Physics, Lomonosov Moscow State University, Moscow, Russia
	}

\shorttitle{Superflare G and K Stars and the Lithium abundance}
\shortauthors{Maria Katsova \&
        Moisey Livshits \&
	Tamara Mishenina \&
	Bulat Nizamov}

\abs{We analyzed here the connection of superflares and the lithium abundance in
 G and K stars based on Li abundance determinations conducted with the echelle spectra
 of a full set of 280 stars obtained with the ELODIE spectrograph. 
 For high-active stars we show a definite correlation between $\log A(Li)$
 and the chromosphere activity. We show that sets of stars with high Li abundance 
 and having superflares possess common properties. It relates, firstly, to stars 
 with activity saturation. We consider the X-ray data for G, K, and M stars separately, 
 and show that transition from a saturation mode to solar-type activity takes place at 
 values of rotation periods 1.1, 3.3, and 7.2 days for G2, K4 and M3 spectral types, 
 respectively. We discuss bimodal distribution of a number of G and K main-sequence 
 stars versus an axial rotation and location of superflare stars with respect to 
 other {\itshape Kepler} stars.  We conclude that superflare G and K stars are mainly fast 
 rotating young objects, but some of them belong to stars with solar-type activity. 
 At the same time, we found a group of G stars with high Li content 
 $(\log A(Li) =  1.5 - 3)$, but being slower rotators with rotation periods > 10 days, 
 which are characterized by low chromospheric activity. This agrees with a large 
 spread in Li abundances in superflare stars. A mechanism leading to this effect 
 is discussed.\\[5pt]
 \textit{Keywords:} \textsl{Stars: activity -- Stars: coronae --
           Stars: rotation -- Stars: flares -- Stars: abundances}\\
 	\textit{E-mail:} \textsl{maria@sai.msu.ru}
 }

\begin{document}

\maketitle

\section{Introduction}

One of the most important results of study of activity of low-mass stars is a 
detection with the \textsl{Kepler} mission of superflares whose total energy in optics 
and near IR-range is $10^{33} - 10^{35}\,$ergs, that is by 2--4 orders of magnitudes 
more than the total energy of the largest flares on the Sun. What distinguishes 
the stars where such cataclisms can occur? This problem is solved in two ways; 
first is determination of fundamental parameters and flare occurrence frequencies. 
This was fulfilled by Maehara et al. (2012) and Notsu et al. (2015a, b) where 
they found that these events are typical for G and partly K stars, which rotate 
faster than field stars and are more magnetically active. Another way is 
comparison of general properties of superflare stars with distinctive features 
of other kinds of active late-type stars. This is comparison of activity at 
different layers of stellar atmospheres. Messina et al. (2003) were the first 
who have analyzed rotational modulation of the optical continuum caused by spots 
and the X-ray fluxes of  stars of different spectral types. Lately new X-ray data 
and observations of optical variability of many late-type stars with the \textsl{Kepler}
misson have appeared. Recently we considered a possibility of occurrence of 
very powerful events on late-type stars in the frameworks of current ideas on 
solar flares (Katsova and Livshits, 2015). 
Wichmann et al. (2014) started a programme to study the nature of the stars 
on which these super flares have been observed. Here we are trying to analyze 
features of those stars, where such extremal phenomena occur, on the base of 
new results of multiwavelength observations of late-type stars.

\section{Where do Superflares Occur?}

As an example, let us consider a restricted set of \textsl{Kepler}'s stars where superflare 
light curves were registered with the higher temporal resolution (1 min). 
We use data for 93 active late-type dwarfs from \textsl{Kepler} short-cadence mode. 
Superflare stars are taken from the list by Balona (2015). We chose the GK stars (there are also a number of late F stars) which are present in (McQuillan et al. 2014). To study their activity, we appealed to data on activity 
indicies of these stars. Photospheric activity is manifested in optical and 
near-IR variability discovered with \textsl{Kepler} mission. Therefore, first of all 
we compare amplitudes of brightness variations with rotation periods of these stars. 
The amplitudes of rotational modulation (ARM) are sometimes referred to as 
$\Delta F/F$ or $R_{\textsl{ppm}}$, and defined by a method by Basri et al. (2011), 
modified by McQuillan et al. (2014). Physically, they are an estimate of 
a minimal spot area, in parts per millions. It is a good estimation when 
the rotation axis of a star is inclined to the line of sight on an angle 
$i = 90\,$degrees. At $i < 20$ degree, this method leads to a large errors 
in evaluations of spot areas.

These ARM values are presented in Fig.1 versus axial rotation periods 
for selected G stars with $T_{\textsl{eff}} = 5043 - 5946\,$K and K stars with 
$T_{\textsl{eff}} =3900 - 5043\,$K separately. Comparison of ARM with the maximal 
spot area on the Sun in the highest cycles (0.3\%\ or ARM up to 3000)
shows that most of superflare stars are characterized by high 
photospheric activity.

\begin{figure}[ht]  
	\centering
	\includegraphics[width=\linewidth]{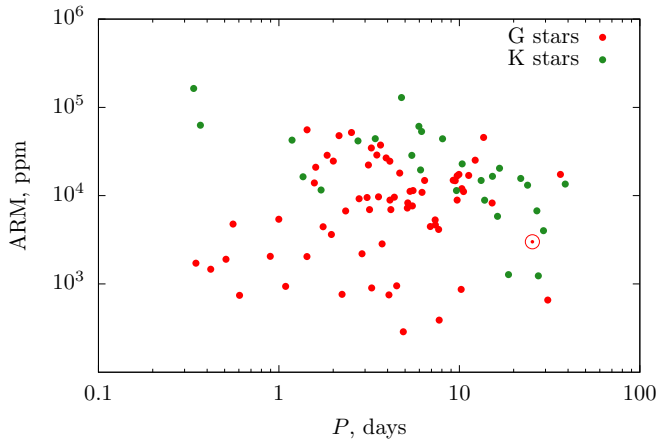}
	\caption{Amplitudes of the rotation modulation (ARM) 
	         for G and late F (red) and K (blue) superflare stars. A dotted empty circle represents the Sun.}
	\label{fig:fig1}
\end{figure}

Note that this is consistent with detailed results by Maehara et al. (2014) 
showing that most of superflare stars rotate with periods of 0.5 -- 7 days 
(a maximum around 5 days). Some of these stars rotate with periods of 
about 15 days.

Here we try to answer a question "Is activity in superflare stars saturated?"
Activity saturation manifests itself the best in the soft X-ray range. 
Research of activity of many of stellar coronae became possible owing to 
the creation of catalog of active F -- M stars based on the X-ray 
observations with space missions \textsl{XMM-Newton}, 
\textsl{ROSAT}, \textsl{Einstein}
(Wright et al. 2011). We plotted again the coronal activity index  
$R_X = \log(L_X /L_{bol})$  versus the rotation period. 
It is reliably shown that there are two groups of stars: 
the first one is a large number of stars with activity saturation 
where $R_X$ is close to $10^{-3}$ and does not depend practically 
on the rotation period. These stars are fast rotators. 
The second group of stars are less active and demonstrate clear 
dependence of $R_X$ on $P_{\textsl{rot}}$. We call activity of stars of 
this group the solar-type activity. For these stars, Skumanich's law, 
$v \propto t^{-1/2}$, where $v$ is axial rotation rate and $t$ is an 
age of a star, is valid. Thus, decline of the coronal activity 
when a star is braking became a basis for a method of gyrochronology 
(age estimate from the activity level). 

Reiners et al. (2014) proposed an original method for an analysis 
of data from the catalog by Wright et al. (2011) and established 
correctly the rotation period, that corresponds to transition 
from saturation to solar-type activity. On the average, this period is 
$P_{sat} = 1.6\,$days.

However all these authors considered all late-type stars together 
without separation onto spectral types. We carried out research, 
similar to that done by Reiners et al., but separately for G, K, and M stars. 
We used the original catalog by Wright et al. (2011) including 824 stars. 
After Reiners et al. (2014), we adopted this relationship as 
$L_X /L_{bol}  = kR^\alpha P^\beta$, where $R$ is a radius of a star
and $P$ is its rotation period.  Details of our approach see in 
Nizamov et al. (2016). Result is given in Fig. 2.  
The transition from a saturation mode to solar-type activity takes place 
at values of rotation periods 1.1, 3.3, and 7.2 days for G2, K4 and M3 
spectral types respectively. Strictly speaking, these values are 
obtained for a case of $L_X \propto P_{\textsl{rot}}^{-2}$.

\begin{figure}[ht]  
	\centering
	\includegraphics[width=\linewidth]{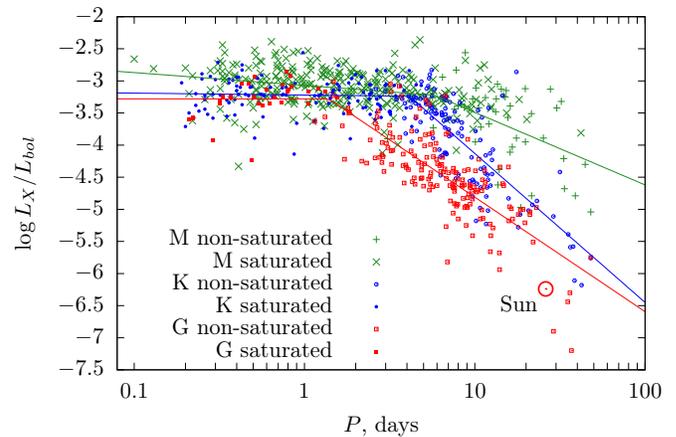}
	\caption{The coronal activity index versus the rotation period}
	\label{fig2}
\end{figure}

In lack of X-ray data for the most of superflare stars, let us consider 
data on the amplitude of rotational modulation, ARM, for them against the 
background of the entire array of \textsl{Kepler} stars. Fig.3 shows location of 
G and K superflare stars (the same as those in Fig.1) among the background 
active \textsl{Kepler} stars, selected from McQuillan et al. (2014) within the 
effective temperature interval $T_{\textsl{eff}} = 5000 - 5500\,$K. It is clearly 
seen that even the minimal level of the photospheric activity of these 
superflare stars exceeds significantly the maximal solar ARM value, 
which is about of 3000. The \textsl{Kepler} data for 34030 stars demonstrate 
distinctly bimodality in distribution of a number of stars over the 
rotation periods (McQuillan et al., 2014). This effect of bimodality is 
the most pronounced for K stars. Separation stars on two subgroups is 
traced both in values ARM and $P_{\textsl{rot}}$. This reflects simultaneous 
existence of stars with activity saturation and with solar-type activity 
both in one range of rotation periods (for instance, 5 -- 15 days) and 
in restricted interval of $T_{\textsl{eff}}$. Bimodality is manifested also 
both in the X-rays and in the $H_\alpha$ data, where there are clearly 
two flux radiation levels for K stars (see Fig.7 in Martinez-Arnaiz et al., 2011). 

Here is important the above defined period corresponding to transition 
from the stars with activity saturation to stars with solar-type activity. 
This period grows from G to M stars. An interval of rotation periods of 
K stars with saturated activity is greater than that of active G stars, 
and these K stars contribute more to the total activity of objects of 
this spectral type. It relates both to quasi-stationary activity and to 
flares including superflares. As for M dwarfs, the character of their 
activity depends on other factors which eliminates this effect.

\begin{figure}[!hb]  
	\centering
	\includegraphics[width=\linewidth]{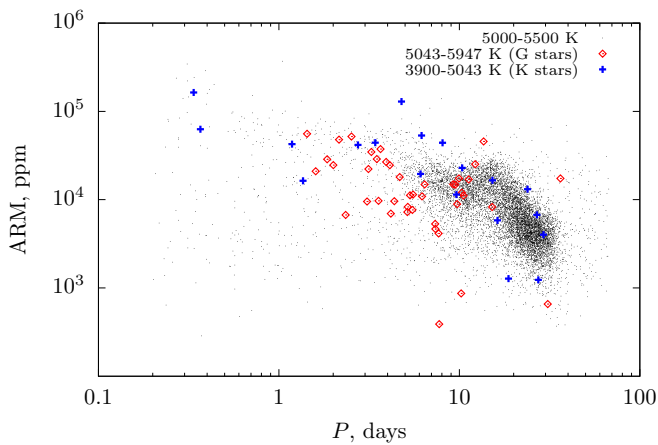}
	\caption{Amplitudes of rotation modulation (or relative spot area) 
	versus $P_{\textsl{rot}}$ for superflare G and K stars against the background 
	other several thousands \textsl{Kepler} main-sequence stars with 
	$T_{\textsl{eff}} = 5000 - 5500\,$K by McQuillan et al. (2014). 
	Note that ARM values < 1000 relate to stars observed under an angle 
	between the rotation axis and the line of sight $i < 20^\circ$.
	}
	\label{fig3}
\end{figure}

Fig.3 shows that superflare stars are situated in the interval of 
rotation periods 1 -- 10 days and values ARM=10000, that is directly 
in the space of activity saturation of G stars. At the same time a 
part of K stars falls into regions of location of slower rotating stars.            

Thus, we conclude that superflare G and K stars are mainly fast rotating 
young objects, but some of them belong to stars with solar-type activity.

\section{The Lithium Abundance in Superflare G and K stars}

It becomes clear that the superflare stars are mainly stars with activity saturation. 
Many of these stars have high lithium abundances. We decide to compare superflare 
stars with those with registered Li 6708$\,\AA$ line. Fig.4 shows the Li abundances 
versus $T_{\textsl{eff}}$~ for more than 1300 late-type dwarfs and subgiants taken from 
Ramirez et al. (2012). 25 superflare stars (blue circles) with the lithium data 
presented by Honda et al. (2015) are plotted against these lithium stars.

\begin{figure}[ht]  
	\centering
	\includegraphics[width=\linewidth]{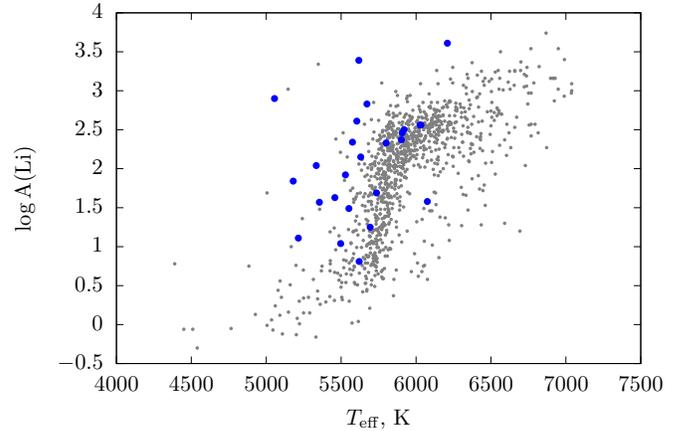}
	\caption{25 Superflare stars (blue circles) from data by Honda et al. (2015) 
	with registered lithium abundances against nearby FGK dwarfs and subgiants 
	(black circles) by Ramirez et al. (2012). Upper limits in the Li abundances 
	of both sets of data are excluded.	
	}
	\label{fig4}
\end{figure}

Here it is seen that superflare stars demonstrate a spread in the lithium abundance, 
and only part of them fall in the same region where most of Li-stars of the same 
spectral type are situated. This relates mostly to K stars, and the high Li abundance 
can be inherent both for K stars with activity saturation and for the case of low-active 
ones (Honda et al. 2015).  Note also that both the lithium content and occurrence of 
superflares are sensitive to the stellar radius. Increase of the radius by 5 -- 10\%\  
leads to changes in the A(Li) and the frequency of occurence and the total energy of 
superflares.

Earlier we found a weak Li abundance-activity correlation for FGK stars which were 
more active than the Sun (Mishenina et al., 2012). These results are based on the 
130 spectra collected with the ELODIE spectrograph using the 1.93-m telescope at 
the Observatoire de Haute Provence. The Li abundances in new set of 150 stars were 
obtained with the same method as Mishenina et al . (2012), by fitting the observational 
profiles to the synthetic spectra  computed with the STARSP LTE spectral synthesis code, 
developed by Tsymbal (1996). A sample of a stellar spectrum in the Li 6708$\,\AA$ line region 
is given in Fig.5. Such a correlation was confirmed on a wider set of more active stars, 
while some inactive stars demonstrated also the high lithium content (Fig. 6 in accordance 
with Katsova et al., 2013). We present here re-analyse of a full set of stars with 
the registered Li line.

\begin{figure}[ht]  
	\centering
	\includegraphics[width=\linewidth]{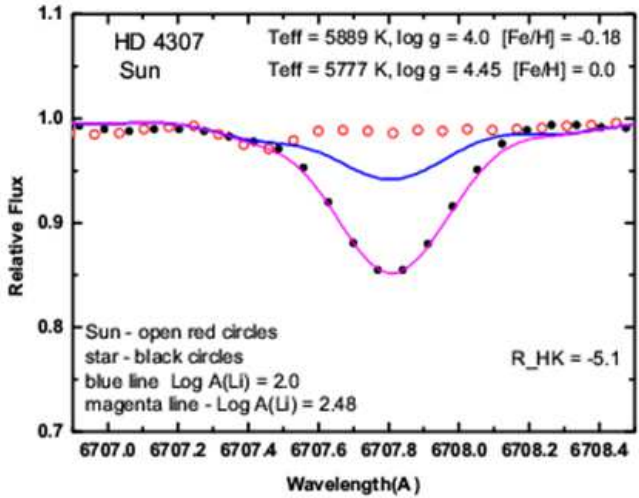}
	\caption{A sample of the spectrum of HD 4307 with well observed Li 6708 \AA\  line
	}
	\label{fig5}
\end{figure}
\begin{figure}[ht]  
	\centering
	\includegraphics[width=\linewidth]{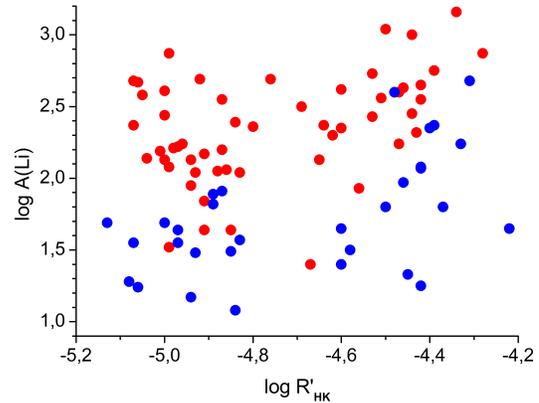}
	\caption{The lithium abundances versus the chromospheric activity index 
		for previous set of stars. Red points are stars hotter than the Sun, 
		blue points present stars cooler than the Sun
	}
	\label{fig6}
\end{figure}

Comparison of both sets presented in Fig. 7 shows that added (new) stars are 
mainly G dwarfs with low chromospheric activity. Note that stars of old and new 
sets have similar metallicity. Results of a new analysis are given in Fig.8a, b and c 
as a dependence of log A(Li) on the effective temperature, rotation velocity $v \sin i$, 
and the index of the chromospheric activity $R^{'}_{\textsl{HK}}$.

\begin{figure}[ht]  
	\centering
	\includegraphics[width=\linewidth]{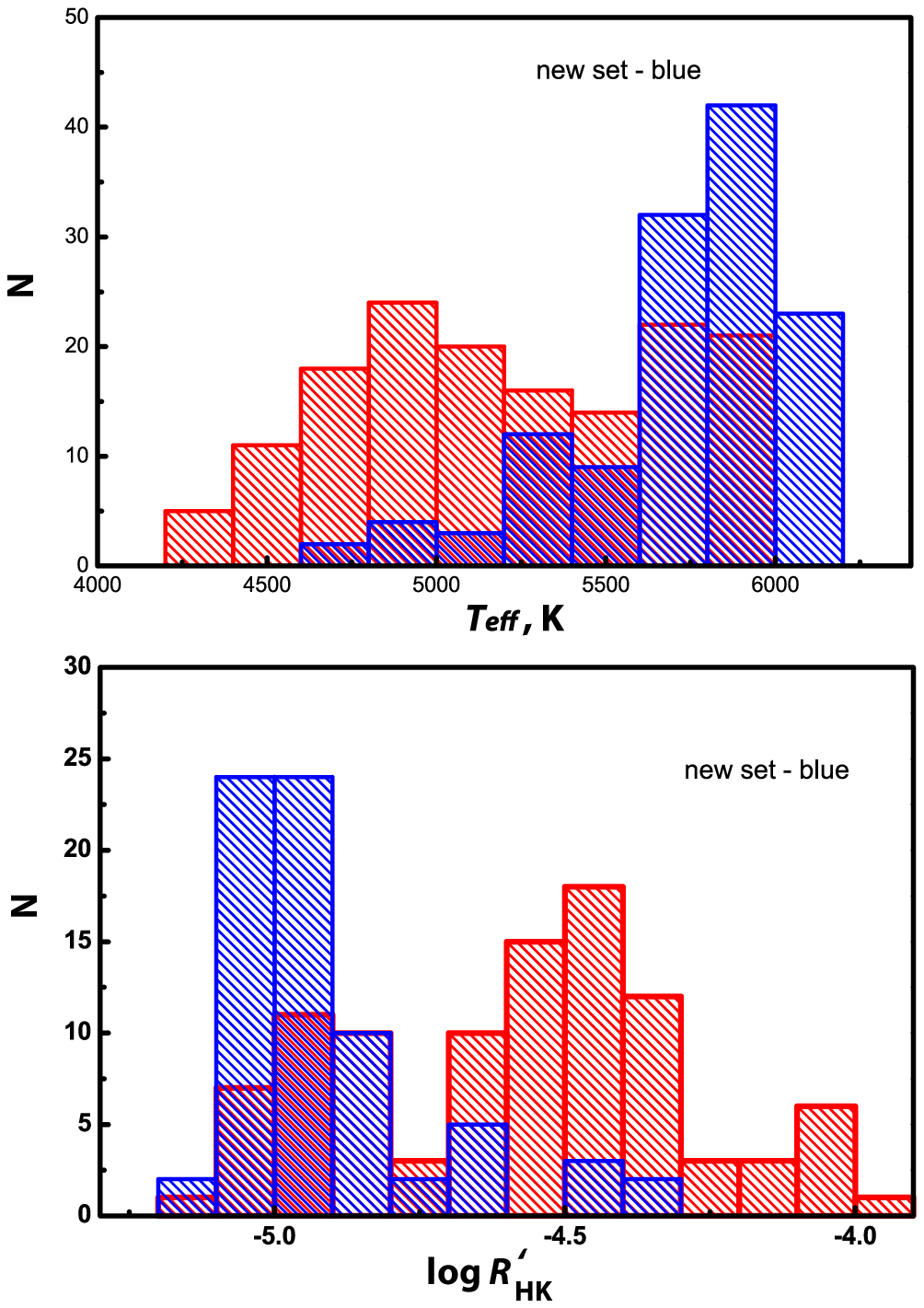}
	\caption{Histograms comparing the previous (red) and new (blue) sets of stars studied
	}
	\label{fig7a}
\end{figure}

This Fig. 8 shows that in fact, we found a group of G stars with high Li content 
($\log A(Li) =   1.5 – 3$), most of which rotate as slow as the Sun. 
These new stars are slower rotators with rotation periods > 10 days.  
Of course, some of new stars are fast rotators with activity saturation. 
The best separation of the slow (blue) and fast (red) rotators is seen in Fig.8c. 
It came as a surprise that high Li-stars can be characterized by the lower 
chromospheric activity,         
$\log R^{'}_{\textsl{HK}} = 10^{-5}$, than the Sun in the minimum of a cycle. 

As it was mentioned above, the part of superflare stars are fast rotators 
with activity saturation. It relates to a wide family of stars with rotation 
periods from a few hours to 8 days, and maximal number of objects rotates with 
period of about 5 days. Many of these stars demonstrate high Li abundances while 
their activity does not depend practically on the rotation period. Our finding 
a group of inactive G stars with high lithium content requires a separate study.

Indeed, the existence of stars with low activity (and hence considered as old) 
but with high lithium abundance is an apparent problem. Why can an old star be 
lithium-rich?

In answer to this question, we suppose  that already on the pre-main sequence 
stage (pMS), solar-like stars may be formed with different lithium abundance. 
After the pMS-stage ends and during MS stage, lithium abundance only slightly 
diminishes in stars more massive than the Sun. 

This picture of lithium evolution is supported by the study of the present Sun. 
Analysis of new helioseismic data show that the rate of lithium depletion is small 
in the convective zone of the present Sun. Therefore the lithium abundance is near 
the same during the Main Sequence stage, and it has decreased from the primordial 
abundance already at early stage.  In assumption of its constant value during the MS-stage, 
the present low solar lithium abundance cannot be explained. So, a conclusion is made 
about crucial role of pMS-stage in the lithium evolution. Observations of solar twins 
also support this hypothesis: there is large dispersion of lithium abundance even in 
young stars and the mean value does not depend on stellar age 
(Oreshina et al. 2015, Thevenin et al. 2016).

In principle, this mechanism can explain a large spread in Li abundances in 
superflare stars reported by Honda et al. (2015). However, such a scenario 
can be realized rather in pMS stars of ages of dozens Myr when initial Li 
abundance does not exceed 1.5 -- 2. In Li-rich stars, rotating much faster, 
processes can occur that lead to decline lithium abundance during life time on MS. 
This problem requires additional consideration.

\section{Conclusions}

The starting point of all our considerations is quite obvious statement that 
superflare stars are characterized by high activity at all the layers of  
the stellar atmosphere. In order to find objective evidence for this assertion, 
we compared firstly activity levels of late-type dwarfs and superflare stars.  
First of all, we came back to problem of coronal activity in stars of different 
spectral types. Here we have continued study of relation between the coronal 
activity index and the rotation period. Earlier Wright et al. (2011) and 
Reiners et al. (2014) conducted an appropriate analysis for all late-type stars. 
We modified an approach proposed by Reiners et al. and obtained results for G, K, 
and M stars, separately. We conclude that transition from a saturation mode 
to solar-type activity takes place at values of rotation periods 1.1, 3.3, and 
7.2 days for G2, K4, and M3 spectral types, respectively. Insofar as most 
superflare stars rotate with periods of 0.5 -- 7 days with a maximum around 5 days, 
as it follows from Maehara et al. (2014), this means that activity of G and K 
superflare stars is saturated or close to saturation mode. This conclusion is 
valid for chromospheres also, as it is seen in IR $Ca\,II$ observations and $H_\alpha$ 
data (Notsu et al. 2015b, Martinez-Arnaiz et al 2011).

Secondly, we compare activity in G and K star photospheres with those 
in superflare stars. \textsl{Kepler} data demonstrate bimodality in distribution 
of a number of active late-type stars. This bimodality presents both in 
rotation periods and in amplitudes of rotation modulation (this means 
apparently bimodality in spot area). On the example of a limited set of 
superflare stars, we have demonstrated similar bimodal distribution in 
a number of stars.     

\begin{figure}[!ht]  
	\centering
	\includegraphics[width=\linewidth]{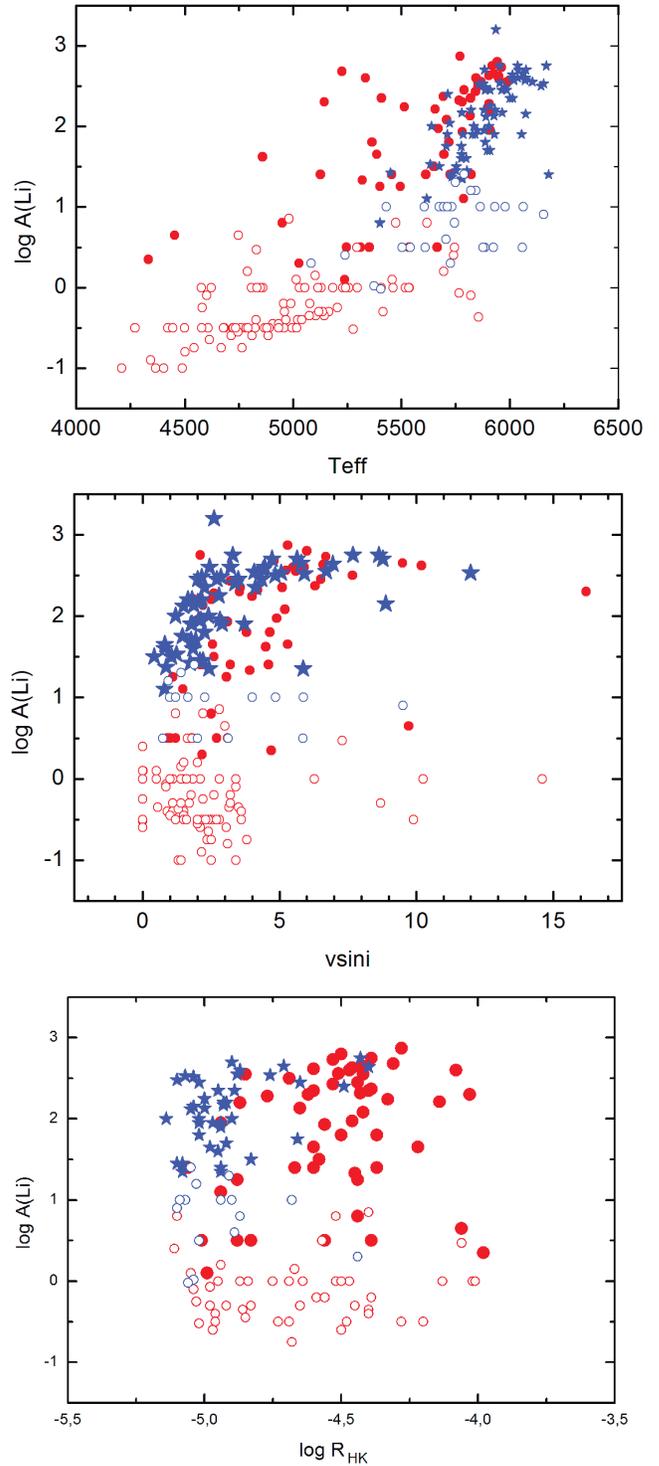}
	\caption{Dependences of log A(Li) on $T_{\textsl{eff}}$ (a), 
	    $\log A(Li)$ on $v \sin i$ (b), and $\log A(Li)$  on $R^{'}_{\textsl{HK}}$ (c).
	    Red colour relates to stars of the pervious set, blue color represents 
		stars of new set. 
        Blue and red open circles are upper limits for log A(Li) in corresponding sets
	}
	\label{fig8}
\end{figure}

Third, involvement of our resuls of study of Li abundances of late-type 
stars supports above mentioned conclusions. Indeed, both superflare stars 
and stars with high Li abundances are characterized by fast rotation and 
are quite young, therefore their activity level and the lithium content 
demonstrate a definite correlation. Besides, we reveal some number of 
G stars with low activity and the high Li abundance.  In these stars, 
processes, responsible for the lithium content, are apparently associated 
with specific initial physical conditions with which they come to the 
main sequence.

\section*{Acknowledgements}

We are thankful to V.~Baturin and A.~Oreshina for fruitful discussion. 
TM  thanks for the support from the Swiss National Science Foundation, 
project SCOPES No.~IZ73Z0152485.
This work is fulfilled in the frameworks of RFBR grants 
14-02-00922 and 15-02-06271, 
and the grant for support of Leading Scientific Schools 9670.2016.2.


\end{document}